\def\year{2022}\relax
\documentclass[letterpaper]{article} % DO NOT CHANGE THIS
\usepackage{aaai22}  % DO NOT CHANGE THIS
\usepackage{times}  % DO NOT CHANGE THIS
\usepackage{helvet}  % DO NOT CHANGE THIS
\usepackage{courier}  % DO NOT CHANGE THIS
\usepackage[hyphens]{url}  % DO NOT CHANGE THIS
\usepackage{graphicx} % DO NOT CHANGE THIS
\usepackage{natbib}  % DO NOT CHANGE THIS AND DO NOT ADD ANY OPTIONS TO IT
\usepackage{caption} % DO NOT CHANGE THIS AND DO NOT ADD ANY OPTIONS TO IT
\DeclareCaptionStyle{ruled}{labelfont=normalfont,labelsep=colon,strut=off} % DO NOT CHANGE THIS
\usepackage{algorithm}
\usepackage{algorithmic}

\usepackage{newfloat}
\usepackage{listings}
\floatstyle{ruled}
\newfloat{listing}{tb}{lst}{}
\floatname{listing}{Listing}
\usepackage{xcolor}
\usepackage{amsmath}
\usepackage{amssymb}
\usepackage{caption}
\usepackage{subcaption}
\usepackage{tikz}
\usepackage{pgfplots}
\usepackage[hidelinks]{hyperref}

\usetikzlibrary{arrows}
\tikzset{auto, >=stealth}
\tikzset{every edge/.append style={shorten >=1pt}}
\pgfplotsset{width=7cm, compat=1.15}
\pgfplotstableset{col sep=comma}
\usepgfplotslibrary{statistics}

\allowdisplaybreaks %SPACING

\usepackage{macros}

\title{
    Learning Temporal Logic Properties: an Overview of Two Recent Methods
}
\author{
	Jean-Rapha\"el Gaglione\textsuperscript{\rmfamily 1},
	Rajarshi Roy\textsuperscript{\rmfamily 2},
	Nasim Baharisangari\textsuperscript{\rmfamily 3},
    Daniel Neider\textsuperscript{\rmfamily 4, \rmfamily 5}\thanks{Part of this work has been conducted when Daniel Neider was at the Carl von Ossietzky University of Oldenburg, Germany},
	Zhe Xu\textsuperscript{\rmfamily 3},
	Ufuk Topcu\textsuperscript{\rmfamily 1}
	\\
}
\affiliations{
	\textsuperscript{\rmfamily 1} University of Texas at Austin, Texas, USA \\
	\textsuperscript{\rmfamily 2} Max Planck Institute for Software Systems, Kaiserslautern, Germany \\
	\textsuperscript{\rmfamily 3} Arizona State University, Arizona, USA \\
	\textsuperscript{\rm 4} TU Dortmund University, Dortmund, Germany\\
	\textsuperscript{\rm 5} Center for Trustworthy Data Science and Security, University Alliance Ruhr, Germany
}

\begin{document}

\maketitle

\begin{abstract}
Learning \emph{linear temporal logic} (LTL) formulas from examples labeled as positive or negative has found applications in inferring descriptions of system behavior.
We summarize two methods to learn LTL formulas from examples in two different problem settings.
The first method assumes noise in the labeling of the examples.
For that, they define the problem of inferring an LTL formula that must be consistent with most but not all of the examples.
The second method considers the other problem of inferring meaningful LTL formulas in the case where only positive examples are given.
Hence, the first method addresses the robustness to noise, and the second method addresses the balance between conciseness and specificity (i.e., language minimality) of the inferred formula.
The summarized methods propose different algorithms to solve the aforementioned problems, as well as to infer other descriptions of temporal properties,
such as \emph{signal temporal logic}
or \emph{deterministic finite automata}.

% \keywords{ % NOT INCLUDED! SEE PDF METADATA INSTEAD (\pdfinfo)
%          Linear Temporal Logic
%     % \and Signal Temporal Logic
%     % \and Decision Tree
%     % \and Specification Mining
%     % \and One-class learning
%     % \and Automata learning
%     \and Learning of logic formulas
%     \and Explainable AI
% }
\end{abstract}

% -- Introduction -------------------------------------------------------------------- %
\section{Introduction}

% \paragraph{Learning properties from examples}
% \textbf{
Learning properties that are interpretable by humans have become a topic of interest with the growing need to explain the behavior of black-box systems.
% }
There is strong motivation to infer such properties from examples when the underlying rules are unknown;
applications range from explaining and debugging to reverse engineering systems that were designed with the help of artificial intelligence.

% \paragraph{Linear Temporal Logic}
% \textbf{
Most systems produce sequences of data, and must be described by properties that capture evolution over time.
% }
\emph{Linear temporal logic} (LTL), amongst other formalisms, models such temporal properties, and expressing properties in LTL makes them particularly easy for humans to interpret.

% \paragraph{Noisy Data}
% \textbf{
\citet{Gaglione2022} assumed noise in the labeling of the examples,
and proposed methods for learning temporal properties that are robust to that noise.
% }
When collecting demonstrations of real-world systems, some examples might be wrongly categorized as positive or negative due to noise in the measurement or uncertainties in the categorization process.
%Naive learning from noisy examples might lead to an inferred formula that does not describe the system accurately.
\citet{Gaglione2022} proposed methods to solve the binary classification problem (i.e., from positive and negative examples) for LTL formulas and for \emph{signal temporal logic} (STL) formulas.

% \paragraph{Positive Data}
% \textbf{
\citet{Roy2022} assumed that only positive examples are available,
and proposed methods for learning concise temporal properties that are specific to the positive examples. 
% }
Negative examples are not always available, and can have a cost to obtain (e.g., crashing a self-driving car or hitting pedestrians).
Learning temporal properties from positive examples only, i.e., the one class classification (OCC) problem, needs to also learn properties that are specific with respect to the positive examples, to avoid trivial solutions such as tautologies. %: adequate properties are properties that have a good balance between conciseness and specificity.
%Language minimality offers a good measure of specificity.
\citet{Roy2022} proposed methods to solve the OCC problem for LTL formulas and for \emph{deterministic finite automata} (DFAs),
using language minimality as a measure for specificity.

% \paragraph{plan}
% \textbf{
We present in this paper the methods and results from both~\cite{Gaglione2022} and~\cite{Roy2022}.
% }

% -- Problem -------------------------------------------------------------------- %
\section{Problem Statements and Methods}

% -- Noisy data ---------------------------------------------------------- %
\subsection{Learning from Noisy Data}
We describe, in Problem~\ref{prob:learn-ltl-imperfect}, the generic form of the problem statement from our first paper \cite{Gaglione2022}.

\begin{problem}[infer an imperfect classification of data]
\label{prob:learn-ltl-imperfect}
Let $\mathcal{S}$ be a labeled sample, i.e.,
a set of traces labeled as positive or negative.
Let be a threshold $\thres \in [0,1]$, expressed as a percentage of noise to tolerate in the sample.
The problem consists in finding an LTL formula $\varphi$ that correctly classifies at least $1-\thres$ of the sample $\mathcal{S}$.
A trace $u$ with label $b \in \{\ltrue,\lfalse\}$ (respectively, positive and negative trace) is considered correctly classified if $\varphi(u) = b$.
\end{problem}

We developed two algorithms to solve Problem~\ref{prob:learn-ltl-imperfect}:
a MaxSAT-based algorithm, referred to as \methNoisy{},
and an algorithm using a decision tree as an extra layer.%, referred to as \methNoisyDT{}.

Note that when $\thres = 0$, Problem~\ref{prob:learn-ltl-imperfect} becomes a perfect classification problem.
This specific case is addressed by \citet{DBLP:conf/fmcad/NeiderG18},
and we used it as a baseline.
We refer to it as \methPerfect{}, since it is not a maximization problem.
%Since the problem becomes a simple SAT problem without maximization, we refer to this baseline as \methPerfect{}.

We also adapted these algorithms to STL formula inference, using \emph{satisfiability modulo theories} (SMT) and MaxSMT instead of SAT and MaxSAT, respectively.
% We refer to this method as \MaxSMT{}.

% -- Positive data ---------------------------------------------------------- %
\subsection{Inference of LTL Formula from Positive Data}

We describe, in Problem~\ref{prob:learn-ltl-specific}, the generic form of the problem statement from \cite{Roy2022}, also referred to as \emph{one class classification} (OCC) problem.

\begin{problem}[infer a language-minimal classification of data]
\label{prob:learn-ltl-specific}
Let $\mathcal{S}$ be a sample composed of (positive) traces.
Let be a bound $\bound\in\N$ for the inferred formula size.
The problem consists in finding an LTL formula $\varphi$ such
that $\abs{\varphi} \leq \bound$,
that it is consistent with the sample $\mathcal{S}$ (i.e., it accepts all the positive traces), and
that it is language-minimal, i.e.,
%for any other LTL formula $\varphi'$ that satisfies the previous constraints, $\varphi \limplies \varphi'$ or $\varphi' \not\rightarrow \varphi$.
there does not exist any other formula $\varphi'$ that satisfies the previous constraints and such that $L(\varphi') \subsetneq L(\varphi)$.
\end{problem}

% We developed three methods to solve Problem \ref{prob:learn-ltl-specific}, namely,
% \methSpecific{CEG}[LTL], a counterexample-guided method based on negative data generation,
% \methSpecific{SYM}[LTL], a symbolic method based on enforcing language strict inclusion during learning, and
% \methSpecific{S-SYM}[LTL], a semi-symbolic method that is a hybrid of the previous ones.

We developed three methods to solve Problem \ref{prob:learn-ltl-specific}, the best of them being referred to as \methSpecific{S-SYM}[LTL].
%
% In \citet{Roy2022},
We also explored similar approaches applied to DFA inference.

% -- Case studies -------------------------------------------------------------------- %
\section{Case Studies}

We evaluated our algorithms on two types of data.
The first type is randomly generated data from arbitrary ground-truth \LTL{} formulas.
The second type of data is data generated by policies learned from reinforcement learning (RL) in the context of DARPA's Competency-Aware Machine Learning (CAML) program,
on a Pusher-robot and an unmanned aerial vehicle (UAV) use cases.
This second type of data have also been clustered in different strategies beforehand, and the motivation of \LTL{} formula inference is to explain the underlying rules of these strategies.
We present here a selection of results from the summarized papers.

%--------------------------------------------------
\subsection{MaxSAT on Common Ground Truths}

\colorlet{colorThres0}{red!90!black}
\colorlet{colorThres5}{blue}
\colorlet{colorThres10}{green!80!purple}
\colorlet{colorMin80}{black}
\colorlet{colorMin60}{black}

We considered 12 arbitrary ground truth \LTL{} formulas, and generated a first set of 148 samples from these formulas.
%Each sample contains from 12 to 1000 traces of length up to 15.
We then created a second set of samples with noise from the first sample, by changing the labels of $5\%$ of the traces.

% Fig:LTL-SIZE
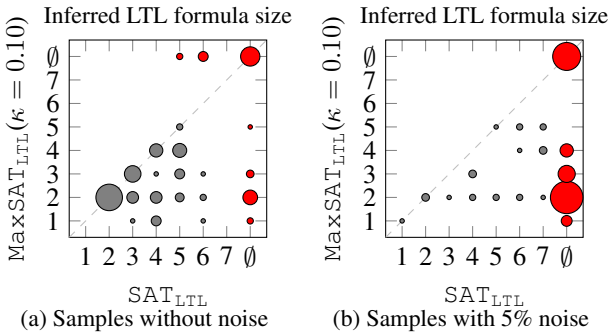
\begin{figure}[ht] %fig:sat-vs-maxsat:size
\centering

\begin{subfigure}{.5\linewidth}
\centering
            % <<<<<<< generated by: pgfgenerate.py bubble --table data-SizeFreq-MaxSATvsSAT-timeouts.csv --columns SAT-perf3:MaxSAT10-perf:Freq-perf3 --title 'Inferred \begin \LTL {} formula size' --xlabel '\SAT {}' --ylabel '\MaxSAT {}($\thres =0.10$)' -o eval-SAT-MaxSAT-perf3.tex
            % required packages: \usepackage{tikz, pgfplots}

            \begin{tikzpicture}
            \begin{axis}
            [
                width = \linewidth,
height = \linewidth,
xmin = 1,
ymin = 1,
xmax = 8,
ymax = 8,
xtick = {1,...,8},
ytick = {1,...,8},
xticklabels = {1,2,3,4,5,6,7,$\emptyset$},
yticklabels = {1,2,3,4,5,6,7,$\emptyset$},
enlarge x limits,
enlarge y limits,
title = {Inferred \LTL {} formula size},
title style = {font=\footnotesize, inner sep=0pt},
xlabel = {\SAT {}},
xlabel near ticks,
ylabel = {\MaxSAT {}($\thres =0.10$)},
ylabel near ticks
            ]

            \addplot[scatter=true,
            	only marks,
            	visualization depends on = {sqrt(\thisrow{Freq-perf3}/148)*0.085*\linewidth \as \perpointmarksize},
            	point meta={TeX code symbolic={
            		\ifthenelse{\( \thisrow{SAT-perf3}>7 \OR \thisrow{MaxSAT10-perf}>7 \)}
            		{\edef\pgfplotspointmeta{timeout}}
            		{\edef\pgfplotspointmeta{normal}}
            	}},
            %	scatter,
            	scatter/classes={
            		normal={fill=gray,/tikz/mark size=\perpointmarksize},
            		timeout={fill=red,/tikz/mark size=\perpointmarksize}
            	},
            %	mark options={fill=gray},
            %	scatter/@pre marker code/.style={/tikz/mark size=\perpointmarksize},
            %	scatter/@post marker code/.style={},
            ] table [
            	x={SAT-perf3},
            	y={MaxSAT10-perf},
                col sep=comma,
            ] {data-SizeFreq-MaxSATvsSAT-timeouts.csv};
            %every mark/.append style={solid, fill=gray},

            % Diagonal
            \draw[black!25,dashed] (rel axis cs:0, 0) -- (rel axis cs:1, 1);

            \end{axis}
            \end{tikzpicture}

            % >>>>>>>
        
\vspace{-3mm}
\caption{Samples without noise}
\label{fig:sat-vs-maxsat:noise0:size}
\end{subfigure}%
\begin{subfigure}{.5\linewidth}
\centering
            % <<<<<<< generated by: pgfgenerate.py bubble --table data-SizeFreq-MaxSATvsSAT-timeouts.csv --columns SAT-noisy3:MaxSAT10-noisy:Freq-noisy3 --title 'Inferred \LTL {} formula size' --xlabel '\SAT {}' --ylabel '\MaxSAT {}($\thres =0.10$)' -o eval-SAT-MaxSAT-noisy3.tex
            % required packages: \usepackage{tikz, pgfplots}

            \begin{tikzpicture}
            \begin{axis}
            [
                width = \linewidth,
height = \linewidth,
xmin = 1,
ymin = 1,
xmax = 8,
ymax = 8,
xtick = {1,...,8},
ytick = {1,...,8},
xticklabels = {1,2,3,4,5,6,7,$\emptyset$},
yticklabels = {1,2,3,4,5,6,7,$\emptyset$},
enlarge x limits,
enlarge y limits,
title = {Inferred \LTL {} formula size},
title style = {font=\footnotesize, inner sep=0pt},
xlabel = {\SAT {}},
xlabel near ticks,
ylabel = {\MaxSAT {}($\thres =0.10$)},
ylabel near ticks
            ]

            \addplot[scatter=true,
            	only marks,
            	visualization depends on = {sqrt(\thisrow{Freq-noisy3}/148)*0.085*\linewidth \as \perpointmarksize},
            	point meta={TeX code symbolic={
            		\ifthenelse{\( \thisrow{SAT-noisy3}>7 \OR \thisrow{MaxSAT10-noisy}>7 \)}
            		{\edef\pgfplotspointmeta{timeout}}
            		{\edef\pgfplotspointmeta{normal}}
            	}},
            %	scatter,
            	scatter/classes={
            		normal={fill=gray,/tikz/mark size=\perpointmarksize},
            		timeout={fill=red,/tikz/mark size=\perpointmarksize}
            	},
            %	mark options={fill=gray},
            %	scatter/@pre marker code/.style={/tikz/mark size=\perpointmarksize},
            %	scatter/@post marker code/.style={},
            ] table [
            	x={SAT-noisy3},
            	y={MaxSAT10-noisy},
                col sep=comma,
            ] {data-SizeFreq-MaxSATvsSAT-timeouts.csv};
            %every mark/.append style={solid, fill=gray},

            % Diagonal
            \draw[black!25,dashed] (rel axis cs:0, 0) -- (rel axis cs:1, 1);

            \end{axis}
            \end{tikzpicture}

            % >>>>>>>
        
\vspace{-3mm}
\caption{Samples with 5\% noise}
\label{fig:sat-vs-maxsat:noise5:size}
\end{subfigure}
%
%\vspace{-6mm}
\caption{
	Inferred \LTL{} formula size comparison of \SAT{} and \MaxSAT{} with threshold $\thres=0.10$ and a timeout of $900s$ on all samples.
	The area of a circle is proportional to the number of samples it represents.
	The timed out instances are represented by $\emptyset$.
}
\label{fig:sat-vs-maxsat:size}
\end{figure}

On the noisy samples, \MaxSAT{} produced results with running time within orders of magnitude less than \SAT{}, and inferred formulas most of the time when \SAT{} timed out, as demonstrated in Figure \ref{fig:sat-vs-maxsat:size}.

%--------------------------------------------------
\subsection{MaxSMT on the Pusher-robot}

\newcommand{\stratA}{strategy 0}
\newcommand{\stratB}{strategy 1}
\newcommand{\stratC}{strategy 2}
\newcommand{\stratD}{strategy 3}

\colorlet{stratA}{green!80!purple}
\colorlet{stratB}{blue}
\colorlet{stratC}{red!80!black}
\colorlet{stratD}{cyan!80!purple}

Four strategies are identified on the Pusher-robot dataset.
For each of these strategies, a sample is created, where 150 positive traces correspond to the considered strategy, and 150 negative traces correspond to other strategies.
We inferred STL formulas from these samples using the \MaxSMT{} method (\MaxSAT{} applied to STL).

% MAXSAT-FLIE RUNTIME
\begin{figure}[ht] %fig:cs2:maxsat
\centering
\begin{tikzpicture}
\begin{semilogyaxis}
[
	width=\linewidth, height=.45\linewidth,
	title = {\MaxSMT{}},
	xlabel = {Maximum misclassification rate threshold $\thres$},
	ylabel = {Running time in $s$},
	legend entries = {\stratA,\stratB,\stratC,\stratD},
	legend style = {font=\scriptsize,legend pos=north east,draw=none},
	title style = {font=\footnotesize, inner sep=0pt},
	x label style = {yshift=1mm},
	xmin=0, xmax=1,
	ymin=0.5, ymax=2000,
    grid=both,
	grid style={line width=.1pt, draw=gray!10},
    major grid style={line width=.2pt,draw=gray!50},
]

\addplot+ [
	const plot
	%jump
	mark left,
	domain=0:1,
	mark=none,
	color = {stratA},
] table [col sep = space,x=misclass,y=runtime] {
	strat LTLsize misclass runtime
	0 9 -1 1e9
	0 4 0.01667 2427.360
	0 3 0.02000 125.431
	0 2 0.02000 4.028
	0 1 0.40333 1.162
	0 0 2 1e0
};

\addplot+ [
	const plot
	%jump
	mark left,
	domain=0:1,
	mark=none,
	color = {stratB},
] table [col sep = space,x=misclass,y=runtime] {
	strat LTLsize misclass runtime
	1 9 -1 1e9
	1 3 0.16000 1353.080
	1 2 0.16000 20.260
	1 1 0.43000 1.328
	1 0 2 1e0
};

\addplot+ [
	const plot
	%jump
	mark left,
	domain=0:1,
	mark=none,
	color = {stratC},
] table [col sep = space,x=misclass,y=runtime] {
	strat LTLsize misclass runtime
	2 9 -1 1e9
	2 3 0.20333 3572.270
	2 2 0.20333 29.531
	2 1 0.48667 1.365
	2 0 2 1e0
};

\addplot+ [
	const plot
	%jump
	mark left,
	domain=0:1,
	mark=none,
	color = {stratD},
] table [col sep = space,x=misclass,y=runtime] {
	strat LTLsize misclass runtime
	3 9 -1 1e9
	3 3 0.15667 2285.500
	3 2 0.19333 37.355
	3 1 0.33333 1.111
	3 0 2 1e0
};

\end{semilogyaxis}
\end{tikzpicture}
\vspace{-6mm}
\caption{
	Impact of the threshold $\thres$ on the running time of \MaxSMT{} for each strategy of the Pusher-robot.
	%Each step corresponds to a certain number of iterations in Algorithm~\ref{alg:max-sat-learner},
	%i.e., to an inferred \STL{} formula of a certain size, with a misclassification rate lower than or equal to $\thres$.
}
\label{fig:cs2:maxsat}
\end{figure}
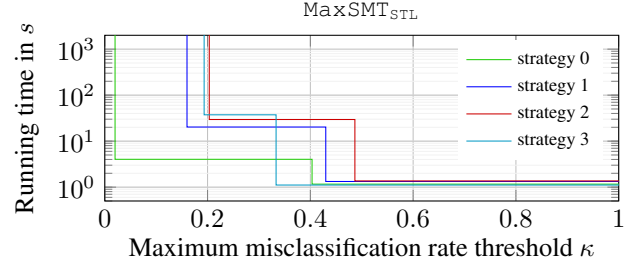

The running time depends on the size of the inferred \STL{} formula, which itself depends on the allowed misclassification rate $\thres$, as presented in Figure~\ref{fig:cs2:maxsat}.
%Figure~\ref{fig:cs2:maxsat} shows the running time of \MaxSMT{} for different numbers of iteration in Algorithm~\ref{alg:max-sat-learner}, presented by misclassification rate.
For example, on the \stratD{} sample,
we could infer the formula $\leventually_{[1,3)} s_0 = 0$
of size 2 with a misclassification rate of $19.33\%$
%(any $\thres \in [0.1933,0.3333)$ would have the same effect), 
and a runtime of 37 seconds.
On the same sample, we could infer the formula
$(s_5 > 0.003) \luntil_{[1,3)} (s_0 = 0)$
of size 3 with a misclassification rate of $15.67\%$
and a runtime of 38 minutes. % (which is way over the chosen timeout but is a good example of non-trivial inferred \STL{} formula).

%--------------------------------------------------
\subsection{One Class Classification on the UAV}

We inferred \LTL{} formulas from three clusters using \methSpecific{S-SYM}[LTL].
The heuristics let us work with very large samples: here, we have a total of 10000 positive traces divided into three samples using $k$-min clustering approach.

\begin{figure}[ht] %fig:UAV
	\begin{center}
	\includegraphics[width=\columnwidth]{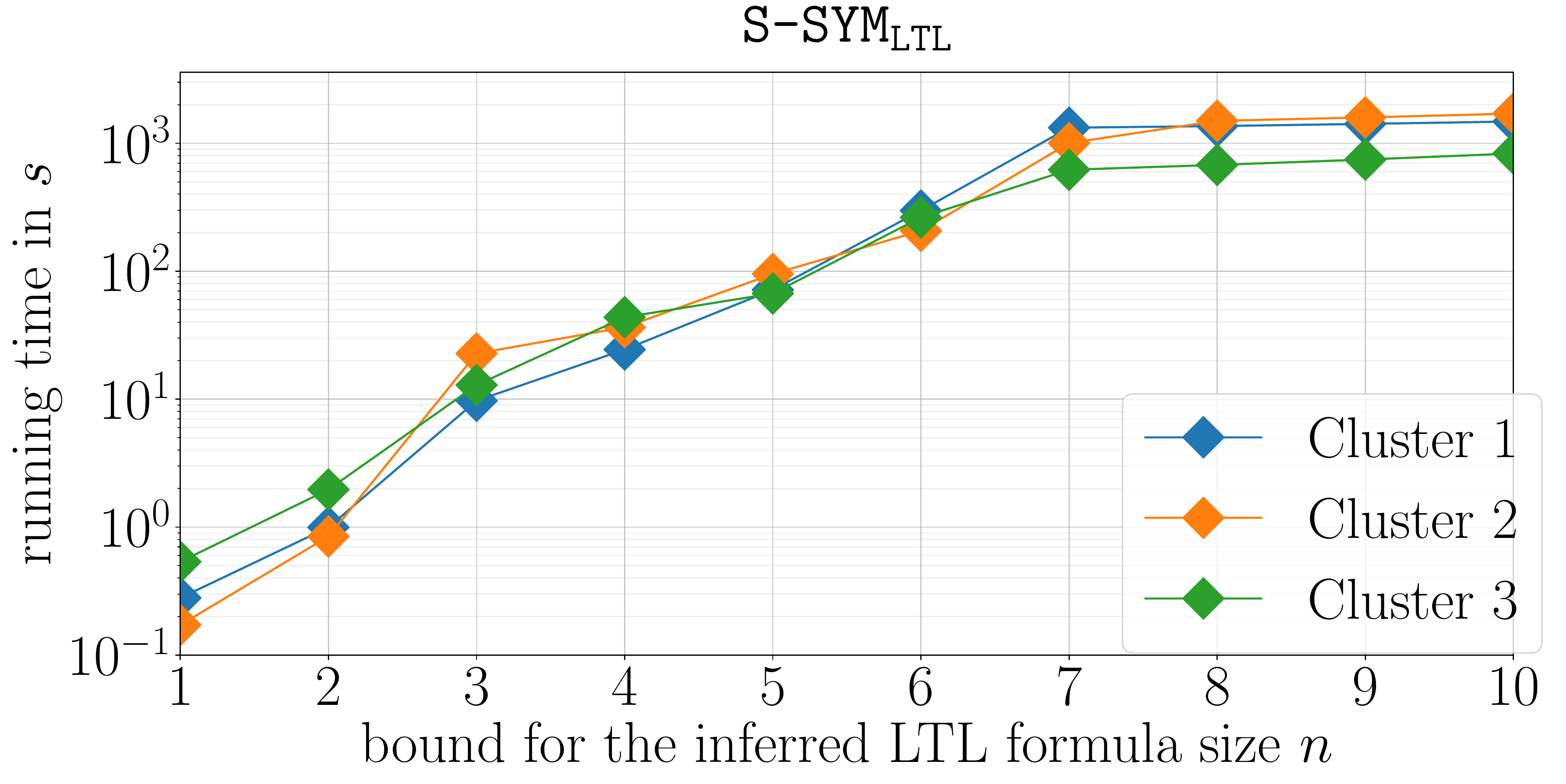}
	\vspace{-6mm}
	\caption{
	    Running time of \methSpecific{S-SYM}[LTL] (OOC problem), for three clusters of traces from a UAV.
	}
	\label{fig:UAV}
	\end{center}
\end{figure}

These three clusters turned out to be not separable, but since we only considered positive data, we still inferred interpretable \LTL{} formulas that ended up characterizing the entire dataset nonetheless.
For example, one of the inferred \LTL{} formula was $(\lF x_3)\limplies(\lG x_3)$ which reads as \textit{``either the UAV always glides, or it never glides''}.
% another one was $\lG(x_8\limplies{x_9})$ which reads as \textit{``a change in yaw angle is always accompanied by a change in roll angle''}.
This formula of size $\bound=4$ was inferred in 36 seconds of running time.
Figure~\ref{fig:UAV} depicts the running time of \methSpecific{S-SYM}[LTL] for different bounds for the formula size.
%On average, \hybridLtl{} is 260.73\% faster than \ceLtl{}.
%Two examples of the inferred \LTL{} formulas from the UAV words are $\lF(x_1)\lor\lnot\lG(x_1)$ which reads as ``\textit{either the UAV always glides, or it never glides}'' and  $\lG(x_2\limplies{x_3})$ which reads as ``\textit{a change in yaw angle is always accompanied by a change in roll angle}''.

\clearpage
\section*{Acknowledgments}
This work has been supported by
the Defense Advanced Research Projects Agency (DARPA) (Contract number HR001120C0032),
Army Research Laboratory (ARL) (Contract number W911NF2020132 and ACC-APG-RTP W911NF),
National Science Foundation (NSF) (Contract number 1646522), and
Deutsche Forschungsgemeinschaft (DFG) (Grant number 434592664).

% -- Bibliography ------------------------------------------------------------ %
\bibliography{bib}

\ifnum\thepage>2
%   \PackageError%
  \PackageWarningNoLine%
  {Guidelines}{\thepage\space pages (should be at most 2 pages)}
\fi

%\iftrue %%%%%%%%%%%%%%%%%%%%%%%%%%%%%%
\iffalse
\clearpage

\section*{\textcolor{magenta}{Goal}}
\url{https://cohrint.info/events/llaama-aaai22/} \\
\url{https://www.aaai.org/Symposia/Fall/fss22.php} \\
2 pages summary (including refs? only summarized papers as refs?) \\
Summary of Noisy Data paper + Positive Data paper

\fi %%%%%%%%%%%%%%%%%%%%%%%%%%%%%%

\end{document}